\documentclass[12pt]{article} 
	 
\usepackage{amsmath} 
\usepackage{psfrag,graphicx,verbatim,array,multicol,palatino,enumerate}
\usepackage{amsmath,amsfonts,bm,rotating,natbib}
\usepackage{pstricks,pst-node}
\usepackage{epsfig}
\usepackage{longtable}
\usepackage[footnotesize,bf]{caption}

\def\balpha{\pmb{\alpha}}
\def\bbeta{\pmb{\beta}}
\def\bgamma{\pmb{\gamma}}

\def\bnu{\pmb{\nu}}
\def\bomega{\pmb{\omega}}

\def\btheta{\pmb{\theta}}
\def\bTheta{\pmb{\Theta}}

\def\bzero{\pmb{0}}
\def\bA{\pmb{A}}   
\def\bB{\pmb{B}}   
   
\def\bE{\pmb{E}}   
\def\bI{\pmb{I}}   
\def\bM{\pmb{M}}

\def\bR{\pmb{R}}   
   
\def\bX{\pmb{X}}   
\def\bY{\pmb{Y}}

\begin{document}

\begin{center}

{\large{\bf HIERARCHICAL BAYESIAN MODELING  OF HITTING PERFORMANCE IN BASEBALL}}

\bigskip

SHANE T. JENSEN\footnote{Department of Statistics, The Wharton School, University of Pennsylvania, \\ Philadelphia, PA 19104  {\tt{stjensen@wharton.upenn.edu}}}, 
BLAKE MCSHANE\footnote{Department of Statistics, The Wharton School, University of Pennsylvania, \\ Philadelphia, PA 19104  {\tt{mcshaneb@wharton.upenn.edu}}} and  
ABRAHAM J. WYNER\footnote{Department of Statistics, The Wharton School, University of Pennsylvania, \\ Philadelphia, PA 19104  {\tt{ajw@wharton.upenn.edu}}}

\end{center}

\bigskip

\begin{abstract}
We have developed a sophisticated statistical model for predicting the
hitting performance of Major League baseball players.  The Bayesian paradigm provides a
principled method for balancing past performance with crucial covariates, such as player age and
position.  We share information across time and across players by using mixture distributions to control shrinkage for improved accuracy.
We compare the performance of our model to current sabermetric methods on a held-out season (2006), and discuss both successes and limitations.  
\end{abstract}

\bigskip

\section{Introduction and Motivation}\label{intro}

There is substantial public and private interest in the projection of future hitting performance in
baseball.  Major league baseball teams award large monetary contracts to top free agent hitters under the assumption that they can reasonably expect  that past success will continue into the future.  Of course, there is an expectation that future performance will vary,  but for the most part it appears that teams are often quite foolishly seduced by a fine performance over a single season. There are many questions:  How should past consistency be balanced with advancing age when projecting future hitting performance?   In young players, how many seasons of above-average performance need to be observed before we consider a player to be a truly exceptional hitter?   What is the effect of a single sub-par year  in an otherwise consistent career? We will attempt to answer these questions through the use of fully parametric statistical models for hitting performance.  

Modeling and prediction of hitting performance is an area of very active research within the baseball community.  Popular current methods include PECOTA \cite[]{Sil03} and MARCEL \cite[]{Tan04b}.   PECOTA is an extremely sophisticated method for prediction of future performance that is based on matching a player's past career performance to the careers of a set of comparable major league ballplayers.  For each player, their set of comparable players  is found by a nearest neighbor analysis of past players (both minor and major league) with similar performance at the same age.  Once a comparison set is found, the future performance prediction for the player is based on the historical performance of those past comparable players.  Factors such as park effects, league effects and physical attributes of the player are also taken into account.  MARCEL is a very simple system for prediction that uses the most recent three seasons of a players major league data, with the most recent seasons weighted more heavily.  There is also a component of the MARCEL methodology that regresses predictions to the overall population mean.   These methods serve as effective tools for prediction, but our focus is on a fully model-based approach that integrates multiple sources of information from publicly available data, such as the Lahman database \cite[]{Lah06}.  One advantage of a model-based approach is the ability to move beyond point predictions to the incorporation of variability via predictive intervals.  
 
In Section~\ref{allmodel}, we present a Bayesian hierarchical model for the evolution of hitting performance throughout the careers of individual players.   Bayesian or Empirical Bayes approaches have recently been used to model individual hitting events based on various within-game covariates \cite[]{QuiMulRos08} and for prediction of within-season performance \cite[]{Bro08}. We are addressing a different question:  how can we predict the course of a particular hitters career based on the seasons of information we have observed thus far?   Our model includes several covariates that are crucial for the accurate prediction of hitting for a particular player in a given year.  A player's age and home ballpark  certainly has an influence on their hitting;  we will include this information among the covariates in our model.    We will also include player position in our model, since we believe that position is an important proxy for hitting performance ({\it e.g.}, second basemen have a generally lower propensity for home runs than first basemen).  Finally, our model will factor past performance of each player into  future predictions.   In Section~\ref{results}, we test our predictions against  a hold out data set, and compare our performance with several competing methods.   As mentioned above, current external methods focus largely on point prediction with little attention given to the variance of these predictions.  We address this issue by 
examining our results not only in terms of accuracy of our point predictions, but also the quality the prediction intervals produced by our model.   We also investigate several other interesting aspects of our model in Section~\ref{results} and then conclude with a brief discussion in Section~\ref{discussion}.  

\bigskip

\section{Model and Implementation}\label{allmodel}

Our data comes from the publicly-available Lahman Baseball Database \cite[]{Lah06}, which contains hitting totals for each major league baseball player from 1871 to the present day, though we will fit our model using only seasons from 1990 to 2005.    In total, we have 10280 player-years of of information from major league baseball between 1990 and 2005 that will be used for model estimation.   Within each season $j$, we will use the following data for each player $i$:
\begin{enumerate}
\item Home Run Total : $Y_{ij}$ 
\item At Bat Total : $M_{ij}$ 
\item Age  :  $A_{ij}$ 
\item Home Ballpark : $B_{ij}$ 
\item Position : $R_{ij}$ 
\end{enumerate}
As an example, Barry Bonds in 2001 had $Y_{ij} = 73$ home-runs out of $M_{ij} = 664$ at bats.   We excluded pitchers from our model, leaving us with nine positions: first basemen (1B), second basemen (2B), third basemen (3B), shortstop (SS), left fielder (LF), center fielder (CF), right fielder (RF), catcher (C), and the designated hitter (DH).  There were 46 different home ballparks used in major league baseball between 1990 and 2005.   Player ages ranged between 20 and 49, though the vast majority of player ages were between 23 and 44. 

\bigskip

\subsection{Hierarchical Model for Hitting}\label{singleeventmodeling}

Our outcome of interest for a given player $i$ in a given year (season)  $j$ is their home-run total $Y_{ij}$, which we model as a Binomial variable:
\begin{eqnarray}
Y_{ij} \sim {\rm Binomial} (M_{ij}, \theta_{ij})  \label{yequation}
\end{eqnarray}
where $\theta_{ij}$ is a player- and year-specific home run rate, and $M_{ij}$ are the number of opportunities (at bats) for player $i$ in year $j$.    Note that by using at-bats as our number of opportunities, we are excluding outcomes such as walks, sacrifice flies and hit-by-pitches.  We will assume that the number of opportunities $M_{ij}$ are fixed and known so we focus our efforts on modeling each home run rate $\theta_{ij}$.   The i.i.d. assumption underlying the binomial model has already been justified for hitting totals within a single season \cite[]{Bro08}, and so seems reasonable for hitting totals across an entire season.  

We next model each unobserved player-year rate $\theta_{ij}$ as a function of home ballpark $b=B_{ij}$, position $k=R_{ij}$ and age $A_{ij}$ of player $i$ in year $j$.  
\begin{eqnarray}
\log \left( \frac{\theta_{ij}}{1-\theta_{ij}} \right) = \alpha_{k}  + \beta_{b} + f_k(A_{ij}) \label{thetaequation}
\end{eqnarray}
 The parameter vector $\balpha = (\alpha_1,\ldots,\alpha_9)$ are the position-specific intercepts for each of the nine player positions.   The function  $f_k(A_{ij})$ is a smooth trajectory of $A_{ij}$, that is different for each position $k$.  We  allow a flexible model for $f_k(\cdot)$ by using a cubic B-spline \cite[]{Boo78} with different spline coefficients $\bgamma$ estimated for each position.     The age trajectory component of this model involves the estimation of 36 parameters: four B-spline coefficients per position $\times$ nine different positions.   

We call the parameter vector $\bbeta$ the ``team effects" since these parameters are shared by all players with the same team and home ballpark.   However, these coefficients $\bbeta$ can not be interpreted as a true ``ballpark effect" since they are confounded with the effect of the team playing in that ballpark.  If a particular team contains many home run hitters, then that can influence the effect of their home ballpark.   Separating the effect of team versus the effect of ballpark would require examining hitting data at the game level instead of the seasonal level we are using for our current model.

There are two additional aspects of hitting performance that are not captured by the model outlined in (\ref{yequation})-(\ref{thetaequation}).   Firstly, conditional on the covariates age, position, and ballpark, our model treats the home run rate $\theta_{ij}$ as independent and identically-distributed across players $i$ and years $j$.  However, we suspect that not all hitters are created equal: we posit that there exists a sub-group of elite home run hitters within each position that share a  higher mean home run rate.  We can represent this belief by placing a mixture model on the intercept term $\alpha_k$ dictated by a latent variable $E_{ij}$ in each player-year.  In other words,  
\[ \alpha_k = \left\{ \begin{array}
           {r@{\qquad {\rm if} \quad}l@{ \quad}l}
           \alpha_{k0} & {\rm E}_{ij} = 0 &\\
          \alpha_{k1} & {\rm E}_{ij} = 1 & \label{mixture} \\ 
           \end{array} \right. \]  

where we force $\alpha_{k0} < \alpha_{k1}$ for each position $k$.  We call the latent variable ${\rm E}_{ij}$ the elite status for player $i$ in year $j$.  Players with elite status are modeled as having the same shape to their age trajectory, but with an extra additive term (on the log-odds scale) that increases their home run rate.   However, we have a different elite indicator ${\rm E}_{ij}$ for each player-year, which means that a particular player $i$ can move in and out of elite status during the course of his career.  Thus, the elite sub-group is maintained in the player population throughout time even though this sub-group will not contain the exact same players from year to year.  

\begin{figure}
\caption{Hidden Markov Model for Elite Status}\label{markovfigure}

\vspace{-0.5cm}

\begin{center}
\includegraphics[height=2in,width=3.5in]{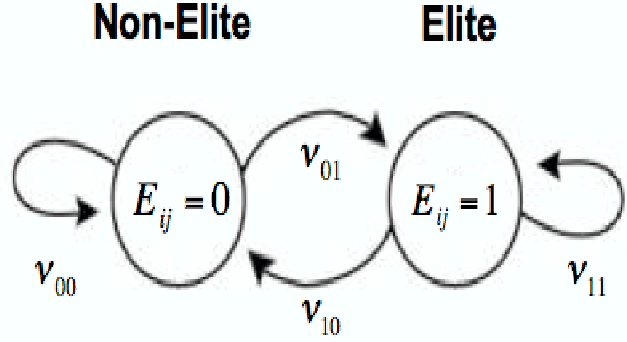}
\end{center}

\vspace{-0.5cm}

\end{figure}

The second aspect of hitting performance that needs to be addressed is that the past performance of a particular player should contain information about his future performance.    One option would be to use player-specific intercepts in the model to allow each player to have a different trajectory.  However, this model choice would involve a large number of parameters, even if these player-specific intercepts were assumed to share a common prior distribution.   In addition, many of these intercepts would be subject to over-fitting due to small number of observed years of data for many players.  We instead favor an approach that involves fewer parameters (to prevent over-fitting) while still allowing different histories for individual players.    We accomplish this goal by building the past performance of each player into our model through a hidden Markov model on the elite status indicators ${\rm E}_{ij}$ for each player $i$.    Specifically, our probability model of the elite status indicator for player $i$ in year $j+1$ is allowed to depend on the elite status indicator for player $i$ in year $j$:
\begin{eqnarray}
p ({\rm E}_{i,j+1} = b | {\rm E}_{ij} = a, R_{ij} = k)  = \nu_{abk} \qquad a,b \in \{0,1\}  \label{Eequation}
\end{eqnarray}
where ${\rm E}_{ij}$ is the elite status indicator and $R_{ij}$ is the position of player $i$ in year $j$.  This relationship is also graphically represented in Figure~\ref{markovfigure}.  The Markovian assumption induces a dependence structure on the home run rates $\theta_{i,j}$ over time for each player $i$.  Players that show elite performance up until year $j$ are more likely to be predicted as elite at year $j+1$.    The transition parameters $\bnu_k = (\nu_{00k},\nu_{01k},\nu_{10k},\nu_{11k})$ for each position $k = 1,\ldots,9$ are shared across players at their position, but can differ between positions, which allows for a different proportion of elite players in each position.  We initialize each player's Markov chain by setting ${\rm E}_{i0} = 0$ for all $i$, meaning that each player starts their career in non-elite status.  This initialization has the desired consequence that young players must show consistently elite performance in multiple years in order to have a high probability of moving to the elite group.  

In order to take a fully Bayesian approach to this problem, we must specify prior distributions for all of our unknown parameters.   The forty-eight different ballpark coefficients $\bbeta$ in our model all share a common Normal distribution, 
\begin{eqnarray}
\beta_{l} & \sim & {\rm Normal} (0, \tau^2) \qquad \forall \quad l = 1, \ldots, 48 \label{priorbeta}
\end{eqnarray}
The spline coefficients $\bgamma$ needed for the modeling of our age trajectories also share a common Normal distribution, 
\begin{eqnarray}
\gamma_{kl} & \sim & {\rm Normal} (0, \tau^2) \qquad \forall \quad k = 1, \ldots, 9, \, l = 1,\ldots,L  \label{priorgamma}
\end{eqnarray}
where $L$ is the number of spline coefficients needed in the modeling of age trajectories for $f(A_{ij},R_{ij})$ for each position.  In our latent mixture model, we also have two intercept coefficients for each position, $\balpha_k = (\alpha_{k0},\alpha_{k1})$, which share a truncated Normal distribution, 
\begin{eqnarray}
\balpha_{k} & \sim & {\rm MVNormal} (\bzero, \tau^2  \bI_2) \cdot {\rm Ind} (\alpha_{k0} < \alpha_{k1}) \qquad \forall \quad k = 1, \ldots, 9 \label{prioralpha}
\end{eqnarray}
where $\bzero$ is the $2 \times 1$ vector of zeros and $\bI_2$ is the $2 \times 2$ identity matrix.   This bivariate distribution is truncated by the indicator function ${\rm Ind} (\cdot)$ to ensure that $\alpha_{k0} < \alpha_{k1}$ for each position $k$.   We make each of the prior distributions (\ref{priorbeta})-(\ref{prioralpha}) non-informative by setting the variance hyperparameter $\tau^2$ to a very large value (10000 in this study).  Finally, for the position-specific transition parameters of our elite status $\bnu$, we use flat Dirichlet prior distributions, 
\begin{eqnarray}
( \nu_{00k}, \nu_{01k} ) & \sim & {\rm Dirichlet} (\omega,\omega) \qquad \forall \quad k = 1, \ldots, 9 \nonumber \\
( \nu_{10k}, \nu_{11k} ) & \sim & {\rm Dirichlet} (\omega,\omega) \qquad \forall \quad k = 1, \ldots, 9
\end{eqnarray}
These prior distributions are made non-informative by setting $\omega$ to a small value ($\omega = 1$ in this study).  We also examined other values for $\omega$ and found that using different values had no influence on our posterior inference, which is to be expected considering the dominance of the data in equation (\ref{gibbstransition}).  Combining these prior distributions together with equations (\ref{yequation})-(\ref{Eequation}) give us the full posterior distribution of our unknown parameters, 
\begin{eqnarray}
p(\balpha,\bbeta,\bgamma,\bnu, \bE | \bX) & \propto & \prod\limits_{i,j} p(Y_{ij} | M_{ij}, \theta_{ij} ) 
\cdot p(\theta_{ij} | R_{ij}, A_{ij}, B_{ij}, E_{ij}, \balpha, \bbeta, \bgamma) \nonumber \\ 
& &\hspace{0.5cm} \cdot p(E_{ij} | E_{i,j-1}, \bnu ) \cdot p(\balpha,\bbeta,\bgamma,\bnu).   \label{fullposterior}
\end{eqnarray}
where we use $\bX$ to denote our entire set of observed data $\bY$ and covariates $(\bA,\bB,\bM,\bR)$.  

\bigskip

\subsection{MCMC Implementation}\label{gibbs}

We estimate our posterior distribution (\ref{fullposterior}) by a Gibbs sampling strategy \cite[]{GemGem84}.  We iteratively sample from the following conditional distributions of each set of parameters given the current values of the other parameters:
\begin{enumerate}
\item $p(\balpha |\bbeta,\bgamma,\bnu, \bE, \bX) = p(\balpha | \bbeta, \bgamma,  \bE, \bX) $
\item $p(\bbeta |\balpha,\bgamma,\bnu, \bE, \bX) = p(\bbeta |\balpha,\bgamma, \bE, \bX) $
\item $p(\bgamma |\bbeta,\balpha,\bnu, \bE, \bX) = p(\bgamma |\bbeta,\balpha, \bE, \bX)$
\item $p(\bnu |\bbeta,\bgamma,\balpha, \bE, \bX) = p(\bnu | \bE) $
\item $p(\bE | \bbeta,\bgamma,\bnu, \bE, \bX) $
\end{enumerate}
where again $\bX$ denotes our entire set of observed data $\bY$ and covariates $(\bA,\bB,\bM,\bR)$.   Combined together, steps 1-3 of the Gibbs sampler represent the usual estimation of regression coefficients ($\balpha,\bbeta,\bgamma$) in a Bayesian logistic regression model.   The conditional posterior distributions for these coefficients are complicated and we employ the common strategy of using the Metropolis-Hastings algorithm to sample each coefficient (see, e.g. \cite{GelCarSte03}).  The proposal distribution for a particular coefficient is a Normal distribution centered at the maximum likelihood estimate of that coefficient.  The variance of this Normal proposal distribution is a tuning parameter that was adaptively adjusted to provide a reasonable rejection/acceptance ratio \cite[]{GelRobGil95}.   Step 4 of the Gibbs sampler involves standard distributions for our transition parameters $\bnu_k = (\nu_{00k},\nu_{01k},\nu_{10k},\nu_{11k})$ for each position $k = 1,\ldots,9$.  The conditional posterior distributions for our transition parameters implied by (\ref{fullposterior}) are
\begin{eqnarray}
(\nu_{00k},\nu_{01k}) |  \bE & \sim & {\rm Dirichlet} \left( {\rm N}_{00k} + \omega, {\rm N}_{01k} + \omega \right) \nonumber \\
(\nu_{11k},\nu_{10k}) |  \bE & \sim & {\rm Dirichlet} \left( {\rm N}_{11k} + \omega, {\rm N}_{10k} + \omega \right) \label{gibbstransition} 
\end{eqnarray}
where ${\rm N}_{abk} = \sum\limits_{i}  \sum\limits_{t=1}^{n_i} {\rm I} ({\rm E}_{i,t} = a, {\rm E}_{i,t+1} = b)$ over all players $i$ in position $k$ and where $n_i$ represents the number of years of observed data for player $i$'s career.  Finally, step 5 of our Gibbs sampler involves sampling the elite status $E_{ij}$ for each year $j$ of player $i$, which can be done using the ``Forward-summing Backward-sampling" algorithm for hidden Markov models \cite[]{Chi96}.   For a particular player $i$, this algorithm ``forward-sums" by recursively calculating 
\begin{eqnarray}
p({\rm E}_{it} | \bX_{i,t}, \bTheta) & \propto & p(X_{i,t} | {\rm E}_{it}, \bTheta) \times p ({\rm E}_{it} | \bX_{i,t-1}, \bTheta) \nonumber \\
  & \propto &  p(X_{i,t} | {\rm E}_{it}, \bTheta) \times \sum\limits_{e = 0}^1 p({\rm E}_{it} | {\rm E}_{i,t-1} = e, \bTheta) \cdot p({\rm E}_{i,t-1} = e | \bX_{i,t-1}, \bTheta) \hspace{1cm}
\end{eqnarray}
for all $t=1,\ldots,n_i$ where $\bX_{i,k}$ represents the observed data for player $i$ up until year $k$,  $X_{i,k}$ represents only the observed data for player $i$ in year $k$, and $\bTheta$ represents all other parameters.  The algorithm then "backward-samples" by sampling the terminal elite state $E_{i,n_i}$ from the distribution $p({\rm E}_{i,n_i} | \bX_{i,n_i}, \Theta)$ and then sampling ${\rm E}_{i,t-1} | {\rm E}_{i,t}$ for $t=n_i$ back to $t=1$.  Repeating this algorithm for each player $i$ gives us a complete sample of our elite statuses $\bE$.     We ran multiple chains from different starting values to evaluate convergence of our Gibbs sampler.  Our results are based on several chains where the first 1000 iterations were discarded as burn-in.  Our chains were also thinned, taking only every eighth iteration, in order to eliminate autocorrelation.  

\subsection{Model Extension: Player-Specific Transition Parameters}\label{pshmmmodeling}

In Section~\ref{singleeventmodeling}, we introduced a hidden Markov model that allows the past performance of each player to influence predictions for future performance.  If we infer player $i$ to have been elite in year $t$ (${\rm E}_{i,t} = 1$), then this inference influences the elite status of that player in his next year, ${\rm E}_{i,t+1}$ through the transition parameters $\bnu_k$.  However, one potential limitation of these transition parameters $\bnu_k$ is that they are shared globally across all players at that position: each player at position $k$ has the same probability of transitioning from elite to non-elite and vice versa.  This model assumption allows us to pool information across players for the estimation of our transition parameters in (\ref{gibbstransition}), but may lead to loss of information if players are truly heterogeneous with respect to the probability of transitioning between elite and non-elite states.  In order to address this possibility, we consider extending our model to allow player-specific transition parameters in our hidden Markov model.  

Our proposed extension, which we call the PSHMM, has player-specific transition parameters $\bnu^i = (\nu^i_{00},\nu^i_{01},\nu^i_{10},\nu^i_{11})$ for each player $i$, that share a common prior distribution, 
\begin{eqnarray}
(\nu^i_{00},\nu^i_{01})  & \sim & {\rm Dirichlet} \left(\omega_{00k} \, , \, \omega_{01k} \right) \nonumber \\
(\nu^i_{11},\nu^i_{10})  & \sim & {\rm Dirichlet} \left(\omega_{11k} \, , \, \omega_{10k} \right)  \label{newprior1} 
\end{eqnarray}
where $k$ is the position of player $i$.  Global parameters $\bomega_k = (\omega_{00k}, \omega_{01k}, \omega_{11k}, \omega_{10k})$ are now allowed to vary with flat prior distributions.    This new hierarchical structure allows for transition probabilities $\bnu^i$ to vary between players, but still imposes some shrinkage towards a common distribution controlled by global parameters $\bomega_k$ that are shared across players with position $k$.   Under this model extension, the new conditional posterior distribution for each $\bnu^i$ is 
\begin{eqnarray}
(\nu^i_{00},\nu^i_{01}) |  \bE & \sim & {\rm Dirichlet} \left( {\rm N}^i_{00} + \omega_{00k} \, , \, {\rm N}^i_{01} + \omega_{01k}  \right) \nonumber \\
(\nu^i_{11},\nu^i_{10}) |  \bE & \sim & {\rm Dirichlet} \left( {\rm N}^i_{11} + \omega_{11k} \, , \, {\rm N}^i_{10} + \omega_{10k}  \right)  \label{newposterior} 
\end{eqnarray}
where ${\rm N}^i_{ab} =  \sum\limits_{t=1}^{n_i-1} {\rm I} (E_{i,t} = a, E_{i,t+1} = b)$.    

To implement this extended model, we must replace step 4 in our Gibbs sampler with a step where we draw $\bnu^i$ from (\ref{newposterior}) for each player $i$.   We must also insert a new step in our Gibbs sampler where we sample the global parameters $\bomega_k$ given our sampled values of all the $\bnu^{i}$ values for players at position $k$.  This added step requires sampling  $(\omega_{00k},\omega_{01k})$ from the following conditional distribution:
 \begin{eqnarray}
 p(\omega_{00k},\omega_{01k} | \bnu) \, \propto \,  \left[ \frac{\Gamma(\omega_{00k} + \omega_{01k})}{\Gamma(\omega_{00k}) \Gamma(\omega_{01k})} \right]^{n_k} \times \left[ \prod\limits_{i = 1}^{n_k}  \nu^i_{00} \right]^{\omega_{00k}-1} \times \left[ \prod\limits_{i = 1}^{n_k}  \nu^i_{01} \right]^{\omega_{01k}-1}  \label{abdist}
\end{eqnarray}
where each product is only over players $i$ at position $k$ and $n_k$ is the number of players at position $k$.   We accomplish this sampling by using a Metropolis-Hastings step with true distribution (\ref{abdist}) and Normal proposal distributions:   $\omega_{00k}^{prop} \sim {\rm N} (\hat{\omega}_{00k}, \sigma^2)$ and $\omega_{01k}^{prop} \sim {\rm N} (\hat{\omega}_{01k}, \sigma^2)$.   The means of these proposal distributions are:
\begin{eqnarray}
\hat{\omega}_{00k} = \overline{\nu}_{00k} \left( \frac{\overline{\nu}_{00k} (1 - \overline{\nu}_{00k})}{s_{0k}^2} - 1 \right)  \quad {\rm and} \quad \hat{\omega}_{01k} = (1-\overline{\nu}_{00k}) \left( \frac{\overline{\nu}_{00k} (1 - \overline{\nu}_{00k})}{s_{0k}^2} - 1 \right)
\end{eqnarray}
with
\begin{eqnarray}
\overline{\nu}_{00k} = \sum\limits_{i=1}^{n_k} \nu^i_{00} \, / \, n_k \qquad {\rm and} \qquad s_{0k}^2 = \sum\limits_{i=1}^{n_k} (\nu^i_{00} -\overline{\nu}_{00k})^2  \, / \, n_k \nonumber
\end{eqnarray}
where each sum is over all players $i$ at position $k$ and $n_k$ is the number of players at position $k$.  
 These estimates $\hat{\omega}_{00k}$ and $\hat{\omega}_{01k}$ were calculated by equating the sample mean $\overline{\nu}_{00k}$ and sample variance $s_{0k}^2$ with the mean and variance of the Dirichlet distribution (\ref{abdist}).   Similarly, we sample $(\omega_{11k},\omega_{10k})$ with the same procedure but with obvious substitutions.

\section{Results and Model Comparison}\label{results}

Our primary interest is the prediction of future hitting events, $Y^\star_{t+1}$ for years $t+1,\ldots$ based on our model and observed data up to year $t$.   We estimate the full posterior distribution (\ref{fullposterior}) and then use
this posterior distribution to predict home run totals $Y^\star_{i,2006}$  for each player $i$ in the 2006 season.  The 2006 season serves as an external validation of our method, since this season is not included in our model fit.   We use our predicted home run totals $\bY^\star_{2006}$ for the 2006 season to compare our performance to several previous methods (Section~\ref{predictions-external}) as well as evaluate several internal model choices (Section~\ref{predictions-internal}).  In Section~\ref{trajectoryresults}, we present inference for other parameters of interest from our model, such as the position-specific age curves.   

\subsection{Prediction of 2006 Home Run Totals: Internal Comparisons} \label{predictions-internal}

We can use our posterior distribution (\ref{fullposterior}) based on data from MLB seasons up to 2005 to calculate the predictive distribution of the 2006 hitting rate $\theta_{i,2006}$ for each player $i$.  
\begin{eqnarray}
p (\theta_{i,2006} | \bX) & = & \int p (\theta_{i,2006} | R_{i,2006}, A_{i,2006}, B_{i,2006}, {\rm E}_{i,2006}, \balpha,\bbeta,\bgamma) \nonumber \\
& & \hspace{0.5cm} \cdot p({\rm E}_{i,2006} | \bE_{i}, \bnu)  p(\balpha,\bbeta,\bgamma,\bnu, \bE_i | \bX) \,
d \balpha \, d \bbeta \, d \bgamma \, d\bnu \, d\bE   \label{predtheta}
\end{eqnarray}
where $\bX$ represents all observed data up to 2005.  This integral is estimated using the sampled values from our posterior distribution $p(\balpha,\bbeta,\bgamma,\bnu, \bE_i | \bX)$ that were generated via our Gibbs sampling strategy.  

We can use the posterior predictive distribution (\ref{predtheta}) of each 2006 home run rate $\theta_{i,2006}$ to calculate  the distribution of the home run total $Y^\star_{i,2006}$ for each player in the 2006 season.  
\begin{eqnarray}
p(Y^\star_{i,2006} | \bX) & = & \int p(Y^\star_{i,2006} | M_{i,2006} , \theta_{i,2006}) \cdot p (\theta_{i,2006} | \bX) \,
d \theta_{i,2006}  \label{predY}
\end{eqnarray}
However, the issue with prediction of home run totals is that we must also consider the number of opportunities $M_{i,2006}$.   Since our overall focus has been on modeling home run rates $\theta_{i,2006}$, we will use the true value of $M_{i,2006}$ for the 2006 season in equation (\ref{predY}).  Using the true value of each $M_{i,2006}$ gives a fair comparison of the rate predictions $\theta_{i,2006}$ for each model choice, since it is a constant scaling factor.  This is not a particularly realistic scenario in a prediction setting since the actual number of opportunities will not be known ahead of time.   

Based on the predictive distribution $p(Y^\star_{i,2006} | \bX)$, we can report either a predictive mean ${\rm E}(Y^\star_{i,2006}  | \bX)$ or a predictive interval $C^\star_i$ such that $p(Y^\star_{i,2006}  \in C^\star_i | \bX) \geq 0.80$.     We can examine the accuracy of our model predictions by comparing to the observed HR totals $Y_{i,2006}$ for the 559 players in the
2006 season, which we did not include in our model fit.   We use the following three comparison metrics:
\begin{enumerate}
\item {\bf RMSE}: root mean square error of predictive means: $\sqrt{\sum\limits_i ({\rm E}(Y^\star_{i,2006} | \bX) - Y_{i,2006})^2 / n}$
\item {\bf Interval Coverage}: fraction of 80\% predictive intervals $C^\star_i$ covering observed $Y_{i,2006}$ 
\item {\bf Interval Width}: average width of 80\% predictive intervals $C^\star_i$
\end{enumerate}

In Table~\ref{internalcomp}, we evaluate our full model outlined in Section~\ref{singleeventmodeling} relative to a couple simpler modeling choices.   Specifically, we examine a simpler version of our model without positional information or the mixture model on the $\alpha$ coefficients.   We see from Table~\ref{internalcomp} that our full model gives proper coverage and a substantially lower RMSE than the version of our model without positional information or the elite/non-elite mixture model.    We also examine a truly simplistic strawman, which is to take last years home run totals as the prediction for this years home run totals (ie. $Y^\star_{i,2006} = Y_{i,2005}$).   Since this strawman is only a point estimate, that comparison is made based solely on the RMSE.   As expected, the relative performance of this strawman model is terrible, with a substantially higher RMSE compared to our full model.  Of course, this simple strawman alternative is rather naive and in Section~\ref{predictions-external}, we compare our performance to more sophisticated external prediction approaches.   

\begin{table}[ht]
\caption{Internal Comparison of Different Model Choices.  Measures are calculated over 559 Players from 2006 season.}\label{internalcomp}
\begin{center}
\begin{tabular}{|l|ccc|}
\hline
 &   & Coverage & Average \\
Model & RMSE & of 80\%& Interval \\
&  & Intervals & Width \\
\hline
Full Model  & 5.30 & 0.855 & 9.81 \\ 
No Position or Elite Indicators & 6.87 & 0.644 & 6.56 \\
Strawman: $Y^\star_{i,2006} = Y_{i,2005}$ & 8.24 & NA & NA \\
Player-Specific Transitions &  5.45 & 0.871 & 10.36 \\ 
\hline
\end{tabular}
\end{center}
\end{table}

We also considered an extended model in Section~\ref{pshmmmodeling} with player-specific transition parameters for the hidden Markov model on elite status, and the validation results from this model are also given in Table~\ref{internalcomp}.   Our motivation for this extension was that allowing player-specific transition parameters might reduce the interval width for players that have displayed consistent past performance.  However, we see that the overall prediction accuracy was not improved with this model extension, suggesting that there is not enough additional information in the personal history of most players to noticeably improve the model predictions.    Somewhat surprisingly, we also see that the width of our 80\% predictive intervals are not actually reduced in this extended model.   The reason is that, even for players with long careers of data, the player-specific transition parameters $\bnu^{i}$ fit by this extended model are not extreme enough to force all sampled elite indicators $E_{i,2006}$ to be either 0 or 1, and so the predictive interval is still wide enough to include both possibilities.     

\subsection{Prediction of 2006 Home Run Totals: External Comparisons} \label{predictions-external}

Similarly to Section~\ref{predictions-internal}, we use hold-out home run data for the 2006 season to evaluate our model predictions compared to the predictions from two external methods, PECOTA \cite[]{Sil03}and MARCEL  \cite[]{Tan04b}, both described in Section~\ref{intro}.  For a reasonable comparison set, we focus our external validation on hitters with an empirical home run rate of least 1 HR every 40 AB in at least one season up to 2005 (minimum of 300 AB in that season).  This restriction reduces our dataset for model fitting down to 118 top home run hitters who all have predictions from the competing methods PECOTA and MARCEL.    As noted above, our predicted home-run totals for 2006 are based on the true number of at bats for 2006.  In order to have a fair comparison to external methods such as PECOTA or MARCEL, we also scale the predictions from these methods by the true number of at bats in 2006.   

Our approach has the advantage of producing the full predictive distribution of future observations (summarized by our  predictive intervals). However, the external method MARCEL does not produce comparable intervals, so we only compare to other approaches in terms of prediction accuracy.  We expand our set of accuracy measures to include not only the root mean square error (RMSE), but also the median absolute error (MAE).  In addition to comparing the predictions from each method using overall error rates, we also calculated ``\% BEST" which is, for each method, the percentage of players for which the predicted home run total $Y^\star_{i,2006}$ is the closest to the true home run total among all methods.   Each of these comparison statistics are given in Table~\ref{externalcomp}.   In addition to giving these validation measures for all 118 players, we also separate our comparison for young players (age $\leq$ 26 years) versus older players (age $>$ 26 years).  

\begin{table}[ht]
\caption{Comparison of our model to two external methods on the 2006 predictions of 118 top home-run hitters.  We also provide this comparison for only young players (age $\leq$ 26 years) versus only older players (age $>$ 26 years).  }\label{externalcomp}
\begin{center}
{\small
\begin{tabular}{|l|ccc|ccc|ccc|}
\hline
Method & \multicolumn{3}{|c|}{{\bf All Players}} &  \multicolumn{3}{|c|}{{\bf Young Players}} &  \multicolumn{3}{|c|}{{\bf Older Players}} \\
  & RMSE &  MAE & \% BEST & RMSE &  MAE & \% BEST & RMSE &  MAE & \% BEST\\
\hline
Our  Model & 7.33 & 4.40 &  41 \% & 2.62 & 1.93 & 62\% & 7.56 & 4.48 & 39\% \\
PECOTA &  7.11 & 4.68 & 28 \% & 4.62 & 3.44 & 0\% & 7.26 & 4.79 & 30\% \\
MARCEL & 7.82 & 4.41 &  31 \% & 4.15 & 2.17 & 38\% & 8.02 & 4.57 & 31\%\\
\hline
\end{tabular}
}
\end{center}
\end{table}

We see from Table~\ref{externalcomp}, that our model is extremely competitive with the external methods PECOTA and MARCEL.    When examining all 118 players, our has the smallest median absolute error and the highest ``\% Best" measure, suggesting that our predictions are superior on these absolute scales.    Our superior performance is even more striking when we examine only the young (age $\leq$ 26 years) players in our dataset.   We have the best prediction on 62\% of all young players, and for these young players, both the RMSE and MAE from our method is substantially lower than either PECOTA or MARCEL.   We credit this superior performance to our sophisticated hierarchical approach that builds in information via position as well as past performance.   

However, our method is not completely dominant: we have a larger root mean square error than PECOTA for older players (and overall), which suggests that our model might be making large errors on a small number of players.     Further investigation shows that our model commits its largest errors for players in the designated hitter (DH) position.   This is somewhat expected, since our model seems to perform best for young players and DH is a position almost always occupied by an older player.   Beyond this, the model appears to be over-shrinking predictions for players in the DH role, perhaps because this player position is rather unique and does not fit our model assumptions as well as the other positions.   However, the validation results are generally very encouraging for our approach compared to previous methods, especially among younger players where a principled balance of positional information with past performance is most advantageous.

\begin{figure}[ht]
\caption{Distribution of number of years of observed data needed to infer elite status (${\rm P}(E_{ij} = 1) \geq 0.5$) among all players determined by our model to be elite during their career.} \label{elitedist}

\vspace{-0.5cm} 

\begin{center}
\includegraphics[width=6in,height=4in]{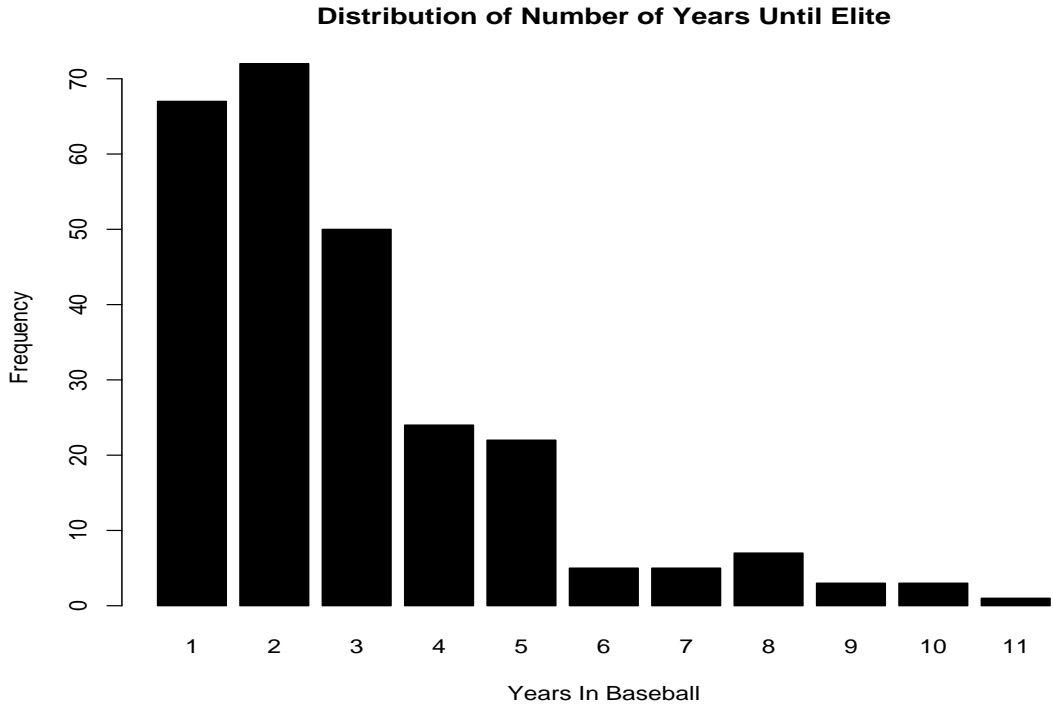}
\end{center}
\end{figure}

We further investigate our model dynamics among young players by examining how many years of observed performance are needed to decide that a player is an elite home run hitter.     This question was posited in Section~\ref{intro} and we now address the question using our elite status indicators ${\rm E}_{ij}$.  Taking all 559 available players examined in Section~\ref{predictions-internal}, we focus our attention on the subset of players that were determined by our model to be in the elite group (${\rm P}(E_{ij} = 1) \geq 0.5$) for at least two years in their career. For each elite home run hitter, we tabulate the number of years of observed data that were needed before they were declared elite.   The distribution of the number of years needed is given in Figure~\ref{elitedist}.   We see that although some players are determined to be elite based on just one year of observed data, most players (74\%) need more than one year of observed performance to determine that they are elite home run hitters.  In fact, almost half of players (46\%) need more than two years of observed performance to determine that they are elite home run hitters.

We also investigated our model dynamics among older players by examining the balancing of past consistency with advancing age, which was also posited as a question in Section~\ref{intro}.   Specifically, for the older players (age $\geq 35$) in our dataset, we examined the differences between the 2006 HR rate predictions $\hat{\theta}_{i,2006} = {\rm E}(\theta_{i,2006} | \bX)$ from our model versus the naive prediction based entirely on the previous year $\tilde{\theta}_{i,2006} = Y_{i,2005}/M_{i,2005}$.   Is our model contribution for a player (which we define as the difference between our model prediction $\hat{\theta}_{i,2006}$ and the naive prediction $\tilde{\theta}_{i,2006}$) more a function of advancing age or past consistency of that player?   Both age and past consistency (measured as the standard deviation of their past home run rates) were found to be equally good predictors of our model contribution, which suggests that both sources of information are being evenly balanced in the predictions produced by our model.

\subsection{Age Trajectory Curves}    \label{trajectoryresults}

\begin{figure}[ht]
\caption{ Age Trajectories $f_k(\cdot)$ for two positions and elite vs. non-elite status.  X-axis is age and Y-axis is Rate =  $\theta_{ij}$} \label{agecurves}


\begin{center}
\includegraphics[width=6.5in,height=5in]{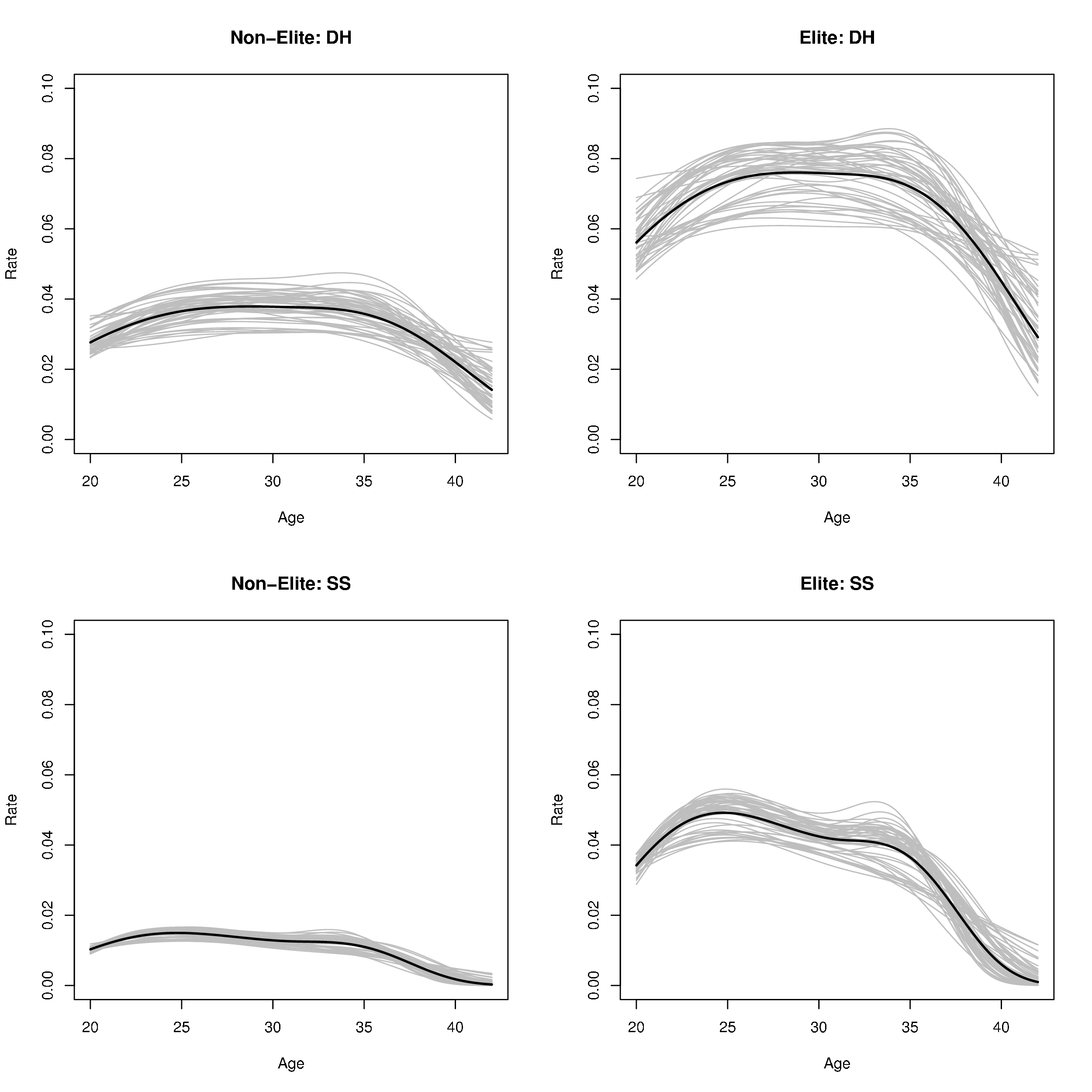}
\end{center}
\end{figure}

In addition to validating our model in terms of prediction accuracy, we can also examine the age trajectory curves that are implied by our estimated posterior distribution (\ref{fullposterior}).  We will examine these curves on the scale of the home run rate $\theta_{ij}$ which is a function of age $A_{ij}$, ball-park $b$, and elite status ${\rm E}_{ij}$ for player $i$ in year $j$ (with position $k$):
\begin{eqnarray}
\theta_{ij}  = \frac{\exp \left[(1-{\rm E}_{ij})\cdot\alpha_{k0} +  {\rm E}_{ij}\cdot\alpha_{k1}  + \beta_{b} + f_k(A_{ij})\right]}{1 + \exp \left[(1-{\rm E}_{ij})\cdot\alpha_{k0} +  {\rm E}_{ij}\cdot\alpha_{k1}   + \beta_b + f_k(A_{ij})\right]} \label{thetaequation2}
\end{eqnarray}
The shape of these curves can differ by position $k$, ballpark $b$ and also can differ between elite and non-elite status as a consequence of having a different additive effect $\alpha_{k0}$ vs. $\alpha_{k1}$.     In Figure~\ref{agecurves}, we compare the age trajectories for two positions, DH and SS, for both elite player-years ($E_{ij} = 1$) vs. non-elite player-years ($E_{ij} = 0$) for an arbitrary ballpark.   Each graph contains multiple curves (100 in each graph), each of which is the curve implied by the sampled values $(\balpha,\bgamma)$ from a single iteration of our converged and thinned Gibbs sampling output.  Examining the curves from multiple samples gives us an indication of the variability in each curve.  

We see a tremendous difference between the two positions DH and SS in terms of the magnitude and shape of their age trajectory curves.  This is not surprising, since home-run hitting ability is known to be quite different between designated hitters and shortstops.  In fact, DH and SS were chosen specifically to illustrate the variability between position with regards to home-run hitting.  For the DH position, we also see that elite vs. non-elite status show a substantial difference in the magnitude of the home-run rate, though the overall shape across age is restricted to be the same by the fact that players of both statuses share the same $f_k(A_{ij})$ in equation (\ref{thetaequation2}).   There is less difference between elite and non-elite status for shortstops, in part due to the lower range of values for shortstops overall.   Not surprisingly, the variability in the curves grows with the magnitude of the home run rate.    

\begin{figure}[ht]
\caption{Distribution of the elite vs. non-elite intercepts  $(\balpha_0,\balpha_1)$ for each position.  The distributions of each  $(\balpha_0,\balpha_1)$ are  presented in terms of the home run rate $\theta_{ij}$ for very young (age = 23) players.  The posterior mean is given as a black dot, and the 95\% posterior interval as a black line. } \label{intercepts}

\vspace{-1.25cm} 

\begin{center}
\rotatebox{270}{\includegraphics[width=4.5in,height=6in]{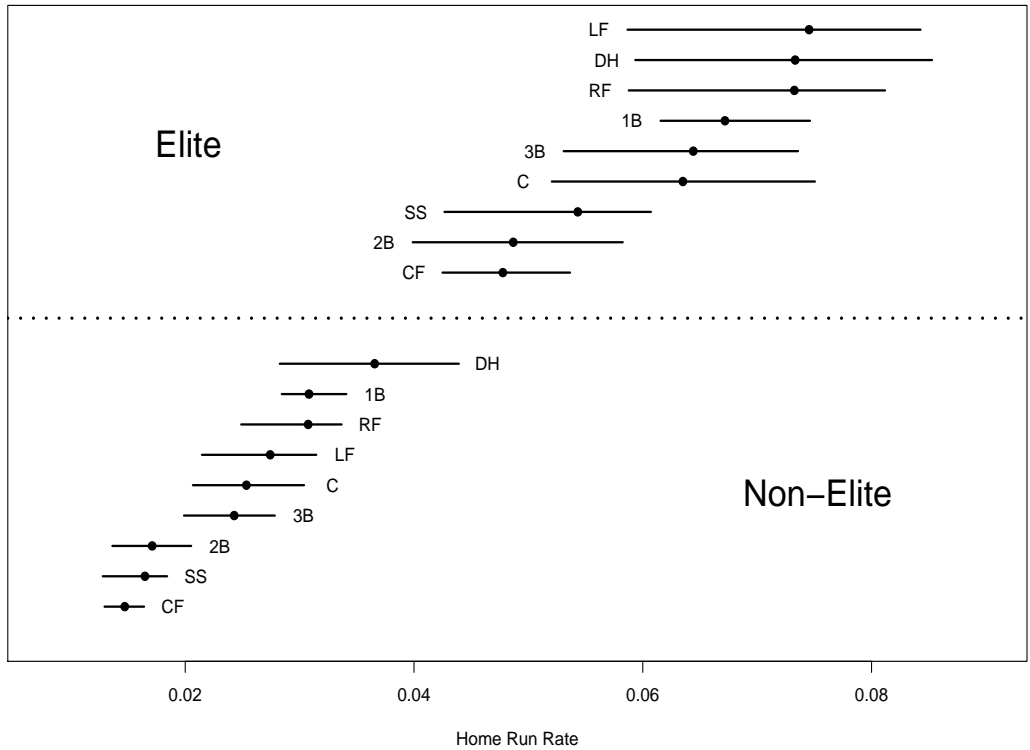}}
\end{center}
\end{figure}

We also perform a comparison across all positions by examining the elite vs. non-elite intercepts $(\balpha_0,\balpha_1)$ that were allowed to vary by position.  We present the posterior distribution of each elite and non-elite intercept  in Figure~\ref{intercepts}.  For easier interpretation, the values of each $\alpha_{k0}$ and $\alpha_{k1}$ have been transformed into the implied home run rate $\theta_{ij}$ for very young (age = 23) players in our dataset.    We see in Figure~\ref{intercepts} that the variability is higher for the elite intercept in each position, and there is even more variability between positions.   The ordering of the positions is not surprising: the corner outfielders and infielders have much higher home run rates than the middle infielder and centerfielder positions.  
 
\begin{figure}[ht]
\caption{Six different hypothetical scenarios for a player at the 2B position.  Black curves indicate the elite and non-elite age trajectories for the 2B position.  Black points represent several several seasons of past performance for a single player.   Predictions for an additional season are given as posterior draws (gray points) of the home run rate and the posterior mean of the home run rate (black $\times$).  Left column of plots gives hypothetical players with consistently high (top row), average (middle row), and poor (bottom row) past home run rates.   Right column of plots show the same hypothetical players, but with their most recent past season replaced by a relatively poor HR hitting performance.} \label{badyear}


\begin{center}
\includegraphics[width=6.5in,height=7.5in]{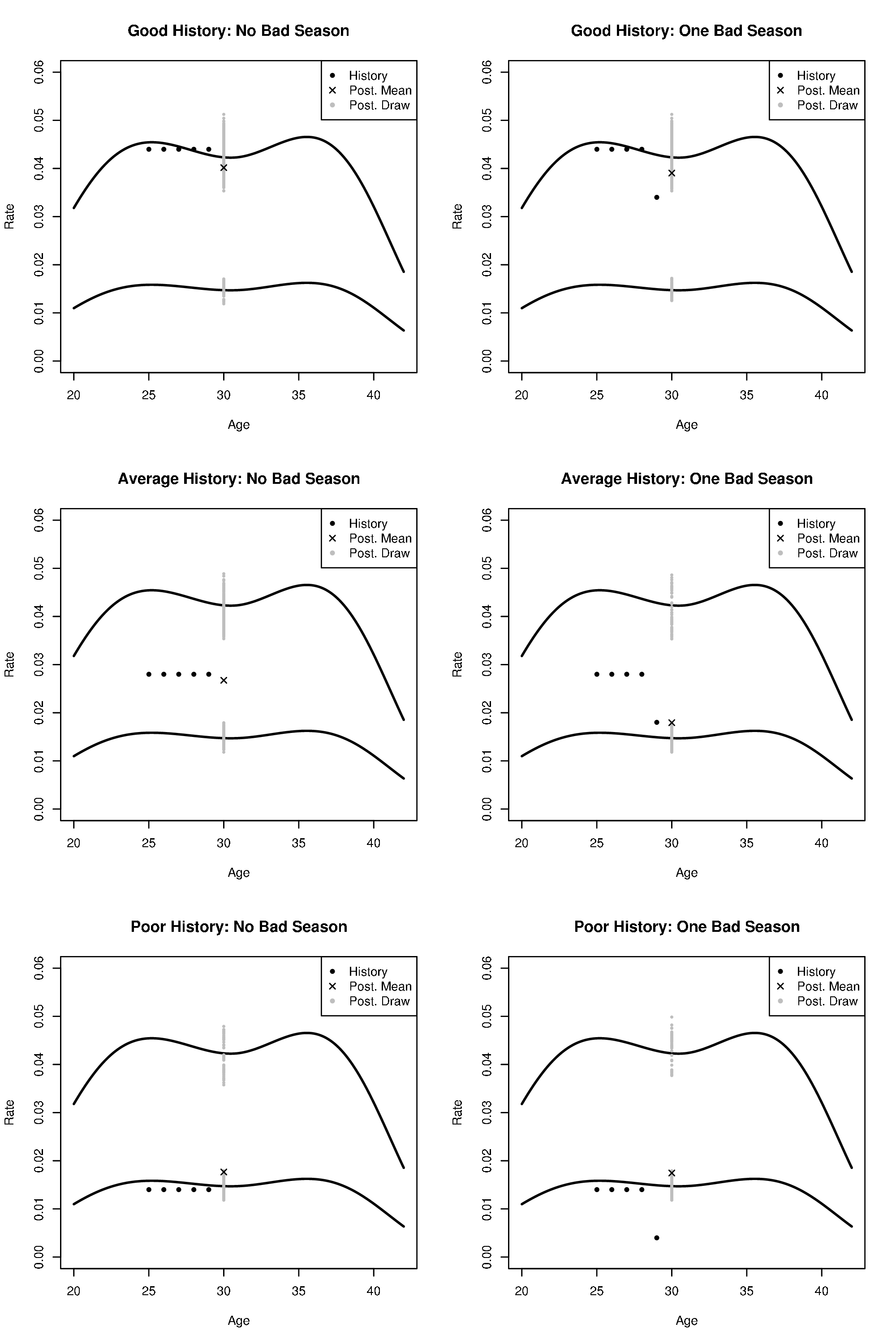}
\end{center}
\end{figure}

For a player at a specific position, such as DH, our predictions of his home run rate for a future season is a weighted mixture of elite and non-elite DH curves given in Figure~\ref{agecurves}.   The amount of weight given to elite vs. non-elite for a given player will be determined by the full posterior distribution (\ref{fullposterior}) as a function of that player's past performance.   We illustrate this characteristic of our model in more detail in Figure~\ref{badyear} by examining six different hypothetical scenarios for players at the 2B position.   Each plot in Figure~\ref{badyear} gives several seasons of past  performance for a single player, as well as predictions for an additional season (age 30).  Predictions are given both in terms of posterior draws of the home run rate as well as the posterior mean of the home run rate.  The elite and non-elite age trajectories for the 2B position are also given in each plot.   We focus first on the left column of plots, which shows hypothetical players with consistently high (top row), average (middle row), and poor (bottom row) past home run rates.   We see in each of these left-hand plots that our posterior draws (gray dots) for the next season are a mixture of posterior samples from the elite and non-elite curves, though each case has a different proportion of elite vs. non-elite, as indicated by the posterior mean of those draws (black $\times$).   

Now, what would happen if each of these players was not so consistent?  In Section~\ref{intro}, we asked about the effect of a single sub-par year on our model predictions.    The plots in the right column show the same three hypothetical players, but with their most recent past season replaced by a season with distinctly different (and relatively poor) HR hitting performance.    We see from the resulting posterior means in each case that only the average player (middle row) has his predictions substantially affected by the one season of relatively poor performance.    Despite the one year of poor performance, the player in the top row of Figure~\ref{badyear} is still considered to be elite in the vast majority of posterior draws.  Similarly, the player in the bottom row of Figure~\ref{badyear} is going to be considered non-elite regardless of that one year of extra poor performance.   The one season of poor performance has the most influence on the player in the middle row, since the model has the most uncertainty with regards to the elite vs. non-elite status of this average player.  

\bigskip

\section{Discussion}\label{discussion}

We have presented a sophisticated Bayesian hierarchical model for home run hitting among major league baseball players.   Our principled approach builds upon information about past performance, age, position, and home ballpark to estimate the underlying home run hitting ability of individual players, while sharing information across players.   Our primary outcome of interest is the prediction of future home run hitting, which we evaluated on a held out season of data (2006).    When compared to the previous methods, PECOTA \cite[]{Sil03} and MARCEL \cite[]{Tan04b}, we perform well in terms of prediction accuracy, especially our ``\% BEST" measure which tabulates the percentage of players for which our predictions are the closest to the truth.  Our approach does especially well among young players, where a principled balance of positional information with past performance seems most helpful.  It is worth noting that our method is fully automated without any post-hoc adjustments, which could possibly have been used to improve the performance of competing methods.    
In addition, our method has the advantage of estimating the full posterior predictive distribution of each player, which provides additional information in the form of posterior intervals.   Beyond our primary goal of prediction, our model-based approach also allows us to answer interesting supplemental questions such as the ones posed in Section~\ref{intro}. 

We have illustrated our methodology using home runs as the hitting event since they are a familiar outcome that most readers can calibrate with their own anecdotal experience.   However, our approach could easily be adapted to other hitting outcomes of interest, such as on-base percentage (rate of hits or walks) which has become a popular tool 
for evaluating overall hitting quality.  Also, although our procedure is presented in the context of predicting a single hitting event, we can also extend our methodology in order to model multiple hitting outcomes simultaneously. In this more general case, 
there are several possible outcomes of an at-bat (out, single, double, etc.).   Our units of observation for a given player $i$ in a given year $j$ is now a vector of outcome totals $\bY_{ij}$, which can be modeled as a multinomial outcome:
$\bY_{ij} \sim {\rm Multinomial} (M_{ij}, \btheta_{ij})$  where $M_{ij}$ are the number of opportunities (at bats) for player $i$ in year $j$ and $\btheta_{ij}$ is the vector of player- and year-specific rates for each outcome.   Our underlying model for the rates $\theta_{ij}$ as a function of position, ball-park and past performance could be extended to a vector of rates $\btheta_{ij}$.   Our preliminary experience with this type of multinomial model indicate that single-event predictions (such as home runs) are not improved by considering multiple outcomes simultaneously, though one could argue that a more honest assessment of the variance in each event would result from acknowledging the possibility of multiple events from each at-bat.  

An important element of our approach was the use of mixture modeling of the player population to further refine our estimated home run rates.  Sophisticated statistical models have been used previously to model the careers of baseball hitters \cite[]{BerReeLar99}, but these approaches have not employed mixtures for the modeling of the player population.    Our internal model comparisons suggest that this mixture model component is crucial for the accuracy of our model, dominating even information about player position.   Using a mixture of elite and non-elite players limits the shrinkage towards the population mean of consistently elite home run hitters, leading to more accurate predictions.    Our fully Bayesian approach also allows us to investigate the dynamics of our elite status indicators directly, as we do in Section~\ref{predictions-external}. 

In addition to our primary goal of home run prediction, our model also estimates several secondary parameters of interest.  We estimate career trajectories for both elite and non-elite players within each position.   In addition to evaluating the dramatic differences between positions in terms of home run trajectories, our fully Bayesian model also has the advantage of estimating the variability in these trajectories, as can be seen in Figure~\ref{agecurves}.     It is worth noting that our age trajectories do not really represent the typical major league baseball career, especially at the higher values of age.  More accurately, our trajectories represent  the typical career conditional on the player staying in baseball, which is one reason why we do not see dramatic dropoff in Figure~\ref{agecurves}.   Since our primary goal is prediction, the fact that our trajectories are conditional is  acceptable, since one would presumably only be interested in prediction for baseball players that are still in the major leagues.  However, if one were more interested in estimating unconditional trajectories, then a more sophisticated modeling of the drop-out/censoring process would be needed.  

Our focus in this paper has been the modeling of home run rates $\theta_{ij}$ and so we have made an assumption throughout our analysis that the number of plate appearances, or opportunities, for each player is a known quantity.   This is a reasonable assumption when retrospectively estimating past performance, but when predicting future hitting performance the number of future opportunities is not known.  In order to maintain a fair comparison between our method and previous approaches for prediction of future totals, we have used the future number of opportunities, which is not a reasonable strategy for real prediction.  A focus of future research is to adapt our sophisticated hierarchical approach to the modeling and prediction of plate appearances $M_{ij}$ in addition to our current modeling of hitting rates $\theta_{ij}$.

\section{Acknowledgements}\label{acknowledgements}
We would like to thank Dylan Small and Larry Brown for helpful discussions.

\bibliography{references}

\end{document}